\documentstyle{mn-1.4}

\newif\ifAMStwofonts
\AMStwofontstrue

\def\sun{\ifmmode\odot\else$\odot$\fi}
\def\H2{\hbox{H$_{2}$}}

\def\kms{${\rm km~s}^{-1}$}

\ifoldfss
  \ifCUPmtlplainloaded \else
    \NewTextAlphabet{textbfit} {cmbxti10} {}
    \NewTextAlphabet{textbfss} {cmssbx10} {}
    \NewMathAlphabet{mathbfit} {cmbxti10} {} 
    \NewMathAlphabet{mathbfss} {cmssbx10} {} 
  \fi
  \ifAMStwofonts
    \ifCUPmtlplainloaded \else
      \NewSymbolFont{upmath} {eurm10}
      \NewSymbolFont{AMSa} {msam10}
      \NewMathSymbol{\upi}     {0}{upmath}{19}
      \NewMathSymbol{\umu}     {0}{upmath}{16}
      \NewMathSymbol{\upartial}{0}{upmath}{40}
      \NewMathSymbol{\leqslant}{3}{AMSa}{36}
      \NewMathSymbol{\geqslant}{3}{AMSa}{3E}

    \fi
  \fi
\fi 

\ifnfssone
  \newmathalphabet{\mathit}
  \addtoversion{normal}{\mathit}{cmr}{m}{it}
  \addtoversion{bold}{\mathit}{cmr}{bx}{it}
  \newmathalphabet{\mathbfit} 
  \addtoversion{normal}{\mathbfit}{cmr}{bx}{it}
  \addtoversion{bold}{\mathbfit}{cmr}{bx}{it}
  \newmathalphabet{\mathbfss} 
  \addtoversion{normal}{\mathbfss}{cmss}{bx}{n}
  \addtoversion{bold}{\mathbfss}{cmss}{bx}{n}
  \ifAMStwofonts
    \ifCUPmtlplainloaded \else
      \UseAMStwoboldmath
      \makeatletter
      \new@mathgroup\upmath@group
      \define@mathgroup\mv@normal\upmath@group{eur}{m}{n}
      \define@mathgroup\mv@bold\upmath@group{eur}{b}{n}
      \edef\UPM{\hexnumber\upmath@group}
      \new@mathgroup\amsa@group
      \define@mathgroup\mv@normal\amsa@group{msa}{m}{n}
      \define@mathgroup\mv@bold\amsa@group{msa}{m}{n}
      \edef\AMSa{\hexnumber\amsa@group}
      \makeatother
      \mathchardef\upi="0\UPM19
      \mathchardef\umu="0\UPM16
      \mathchardef\upartial="0\UPM40
      \mathchardef\leqslant="3\AMSa36
      \mathchardef\geqslant="3\AMSa3E
    \fi
  \fi
\fi 

\ifnfsstwo
  \DeclareMathAlphabet{\mathbfit}{OT1}{cmr}{bx}{it}
  \SetMathAlphabet\mathbfit{bold}{OT1}{cmr}{bx}{it}
  \DeclareMathAlphabet{\mathbfss}{OT1}{cmss}{bx}{n}
  \SetMathAlphabet\mathbfss{bold}{OT1}{cmss}{bx}{n}
  \ifAMStwofonts
    \ifCUPmtlplainloaded \else
      \DeclareSymbolFont{UPM}{U}{eur}{m}{n}
      \SetSymbolFont{UPM}{bold}{U}{eur}{b}{n}
      \DeclareSymbolFont{AMSa}{U}{msa}{m}{n}
      \DeclareMathSymbol{\upi}{0}{UPM}{"19}
      \DeclareMathSymbol{\umu}{0}{UPM}{"16}
      \DeclareMathSymbol{\upartial}{0}{UPM}{"40}
      \DeclareMathSymbol{\leqslant}{3}{AMSa}{"36}
      \DeclareMathSymbol{\geqslant}{3}{AMSa}{"3E}
    \fi
  \fi
\fi 

\ifCUPmtlplainloaded \else
  \ifAMStwofonts \else 
    \def\upi{\pi}
    \def\umu{\mu}
    \def\upartial{\partial}
  \fi
\fi

\title[Circumnuclear star formation in M100]
       {High-resolution {\em UKIRT\/} observations of
        circumnuclear star formation in M100}
\author[S.~Ryder and J.~Knapen]
       {S. D. Ryder$^{1}$ and J. H. Knapen$^{2}$\\
        $^{1}$ Joint Astronomy Centre, 660 N. A'Ohoku Place, Hilo, HI 96720, 
        U.S.A. E-mail: sryder@jach.hawaii.edu.\\
        $^{2}$ Department of Physical Sciences, University of Hertfordshire,
        Hatfield, Herts AL10 9AB, U.K.
        E-mail: knapen@star.herts.ac.uk.}
        \date{Accepted ???.
      Received ???;
      in original form ???}

\pagerange{\pageref{firstpage}--10}
\pubyear{1998}

\begin{document}

\maketitle

\label{firstpage}

\begin{abstract}

We present high-resolution, near-infrared imaging of the circumnuclear
region of the barred spiral galaxy M100 (=NGC~4321), accompanied by
near-infrared spectroscopy. We identify a total of 43~distinct regions
in the $K$-band image, and determine magnitudes and colours for 41
of them. By comparison with other near-infrared maps we also derive
colour excesses and $K$-band extinctions for the knots. Combining the
imaging and spectroscopic results, we conclude that the knots are the
result of bursts of star formation within the last 15--25~Myr. We
discuss the implications of these new results for our dynamical and
evolutionary understanding of this galaxy.

\end{abstract}


\section[]{Introduction}

The circumnuclear regions (CNRs) of barred spiral galaxies, with or
without an active galactic nucleus, often show significant star forming
activity, frequently organized into ring-like structures (e.g., Morgan
1958; S\'{e}rsic \& Pastoriza 1967; Pogge 1989a, b; Buta \& Combes
1996). The massive star formation (SF) is usually observed by imaging in
the H$\alpha$ emission line, and is understood to occur as a result of
orbit crowding and gas accumulation between a pair of dynamical Inner
Lindblad Resonances (ILRs) in the disc. Although the SF in the
circumnuclear `rings' can be explained by gravitational instabilities
there (Elmegreen 1994), spiral structure is almost always present,
either in the SF itself (e.g., M100: Pogge 1989a; Cepa \& Beckman 1990;
Knapen et al. 1995a, hereafter Paper~I), or in dust lanes associated
with the CNR (e.g., NGC~5248: Laine et al. 1998), indicating that the SF
is organized, e.g., by density waves. Near-infrared (NIR) imaging of
CNRs gives important clues on the distribution of obscuring dust
(Paper~I), and can also be used to derive the distribution of stellar
mass, which in turn is used as input for dynamical modelling (e.g.,
Quillen, Frogel \& Gonzalez 1994; Wada, Sakamoto \& Minezaki 1998;
Garc\'{\i}a-Burillo et al. 1998). The NIR light distribution is not used
as input to models by e.g., Knapen et al. (1995b; hereafter Paper~II),
Wada et al. (1998), and Yuan \& Kuo (1998), but these authors do compare
the output of their models with the observed NIR morphology. In Paper~II,
we estimated the value of the $K$-band mass-to-luminosity ratio
($M/L_K$) in the CNR of M100, concluding that $M/L_K$ is not constant,
and is strongly affected by SF. Wada et al. (1998) subsequently
confirmed this from a comparison of their two models.

The main interpretational issue in this respect is what fraction of the
NIR emission traces the stellar mass distribution, and what fraction is
from dynamically young stars? In Paper~II this issue was discussed
qualitatively in detail, concluding that in the CNR of M100 the latter
fraction was significant. Whereas the $K$ light is distributed smoothly
as compared to blue light or H$\alpha$ in M100 (cf. e.g. NGC~4303:
Elmegreen et al. 1997), Knapen et al. (1995b) drew attention to two $K$-band
`hot spots', named K1 and K2, which were identified as starburst regions
at the loci where a pair of trailing and leading shocks interact in the
vicinity of the inner ILR.  Wozniak et al. (1998) subsequently showed
that at 6.75 and 15~$\mu$m, K1 and K2 emit similarly, as expected in
such a scenario. Knapen et al. also reported the detection of `leading
arms' in $K$, in the inner bar region, predicted theoretically.

In this paper, we quantitatively confirm previous suspicions
concerning the stellar populations in the CNR of M100, based on an
analysis of high-resolution NIR imaging of the CNR, and of NIR
spectroscopy of K1, K2, and the nucleus. After describing the
observations (Sect.~\ref{s:obs}) and results (Sect.~\ref{s:res}), we
discuss the ages of the compact knots discovered from the NIR imaging,
and the stellar populations within them. In Sect.~\ref{s:disc}, we
also discuss how the new results fit in with previous work on the CNRs
in M100 and other galaxies. We conclude briefly in Sect.~\ref{s:conc}.

\section[]{Observations}\label{s:obs}

\subsection[]{High-resolution imaging}\label{s:ircam}

The nuclear region of M100 was imaged with IRCAM3 on the upgraded
3.8-m United Kingdom Infrared Telescope ({\em UKIRT}\/) and a (warm)
magnifier, which yielded a pixel scale of $0.140$~arcsec and a
36~arcsec square field of view. Images in $J$, $H$, and $K$ were
obtained on the night of 1998~February~14 UT; bracketing observations
of PSF stars yielded seeing FWHM of 0.34~arcsec at $K$ and 0.39~arcsec
at $J$. For each filter, a $3\times3$ mosaic of overlapping frames
spaced 4~arcsec apart was obtained on M100, together with a matching
mosaic on the sky 10~arcmin away. The integration time per frame was
60~sec. After alignment, the frames in each mosaic were combined,
excluding known bad pixels, and the sky mosaics were used to flatfield
the object mosaics. Owing to their small field coverage, the image
mosaics had to be flux calibrated using aperture photometry from the
literature (Devereux 1989; Recillas-Cruz et al. 1991).

\subsection[]{Near-infrared spectroscopy}\label{s:cgs4}

Longslit $H$ and $K$ band spectra were obtained separately on the
night of 1998~February~12 UT using CGS4 on {\em UKIRT}. A slit width
of 1.2~arcsec was used with the 40~line~mm$^{-1}$ grating in first
order, yielding resolving powers (after smoothing) at $H$ and $K$ of
285 and 360, respectively. The slit position angle was set to
$114^{\circ}$, allowing us to capture both the $K$-band hotspots K1
and K2 identified in Paper~I, as well as the nucleus in the same
observation. Since the infrared-emitting central region of M100 fills
a sizable fraction of the 90~arcsec slit length, sky spectra were
taken with the object positioned well off the slit, rather than simply
displaced along the slit. Total on-source integration times were
1920~sec and 2400~sec at $H$ and $K$, respectively. Observations of
the A4~V star BS~4632 and the F0~V star BS~4531 were used to divide
out atmospheric absorption features in the M100 spectra, after first
having their own Brackett and Paschen lines removed. Preliminary image
processing (subtraction of object--sky pairs and flatfielding) was
performed using the on-line reduction software {\sc cgs4dr}, while
object extraction, wavelength calibration, and division by the
standard were done with {\sc iraf}\footnote{IRAF is distributed by the
National Optical Astronomy Observatories, which are operated by the
Association of Universities for Research in Astronomy, Inc., under
cooperative agreement with the National Science Foundation.}.

\section[]{Results}\label{s:res}

\subsection[]{Photometry}\label{s:phot}

\begin{figure*}
\vspace{18cm}
\includegraphics{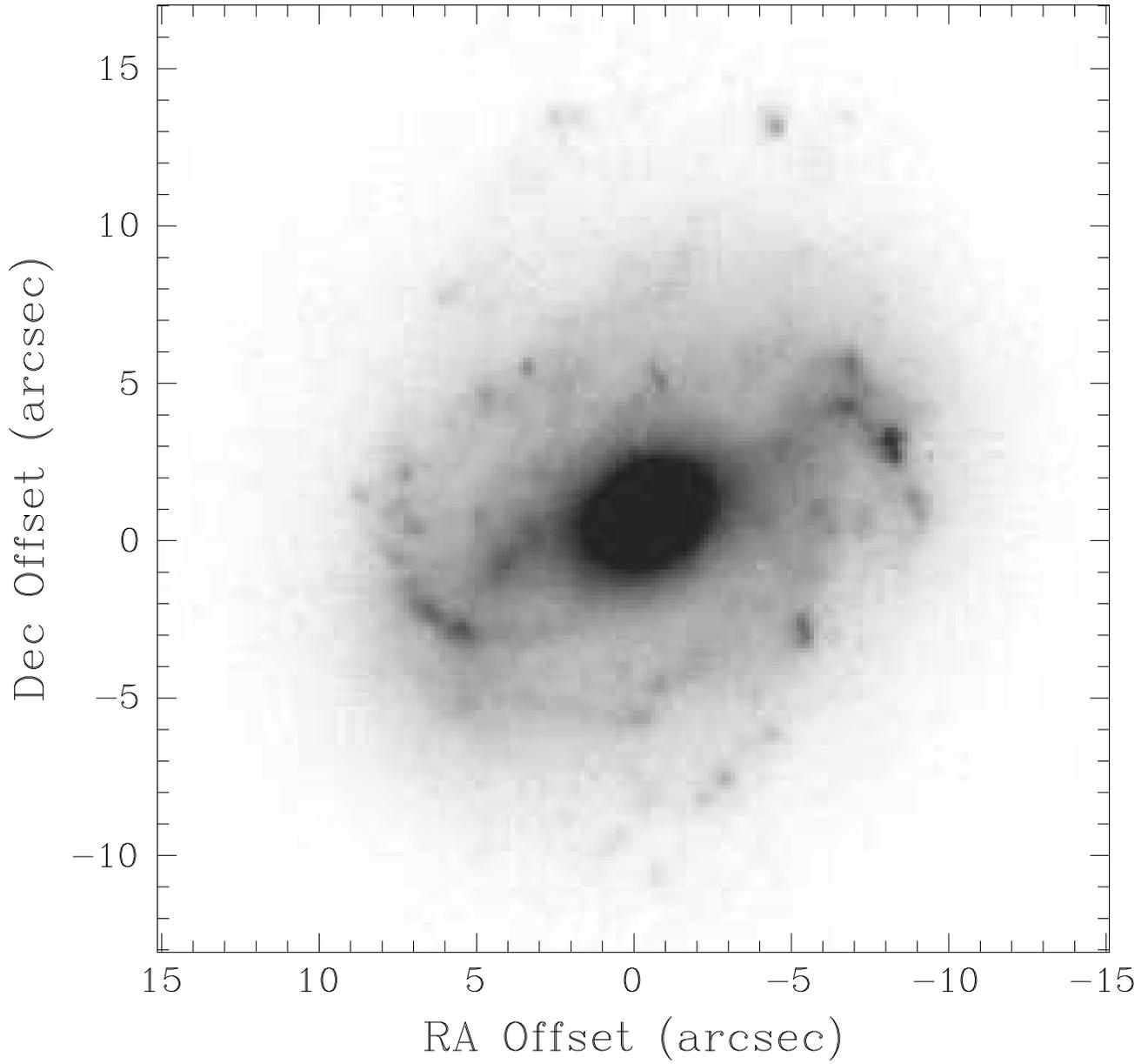}
\caption{$K$-band image of the nuclear region of M100, taken with IRCAM3 and a
$2\times$ magnifier on {\em UKIRT\/}. Offsets are given relative to the
$K$-band nucleus position.}
\label{f:kmap}
\end{figure*}

Figure~\ref{f:kmap} shows the reduced $K$-band image. A total of
43~compact `knots' were identified by a comparison of this image with
an unsharp-masked version (cf. Mighell \& Rich 1995). The improvement
in delivered image quality as a result of the {\em UKIRT\/} Upgrades
Program (which includes active control of the primary mirror figure
and telescope focus, a tip-tilt Fast Guider, and dome ventilation
louvres; Hawarden et al. 1998) can be clearly seen by comparing this
image with the pre-upgrade {\em UKIRT\/} image of Paper~I (their
fig.~2). What were previously just broad maxima in the $K$-band light
have now been resolved into a host of individual regions. For example,
both the maxima K1 and K2 identified in Paper~I at 0.8~arcsec
resolution turn out to consist of 3 or 4 sub-clumps at 0.4~arcsec
resolution. The detection of leading arms within the SF `ring'
(Paper~I) is confirmed on the new images.


Photometry of each of these knots in all 3 bands was performed using
the {\sc ccdcap} digital circular aperture photometry code developed
by Mighell to analyze {\em HST\/} WF/PC and WFPC2 observations
(Mighell 1997, and references therein). The local (sky + unresolved
bulge light) background was determined from a robust estimate of the
mean intensity value of all pixels between 15 and 20 pixels of the
aperture centre. To minimise the effects of crowding, magnitudes were
determined by summing the flux within a radius of 5~pixels
(0.7~arcsec), then corrected to an aperture radius of 15~pixels based
on similar measurements of the PSF stars. The positions,
$K$-magnitudes, and NIR colours (uncorrected for reddening), colour
excesses and inferred $K$-band extinctions (see below) for the 41
knots that could be reliably measured are tabulated in
Table~\ref{t:knots}.

As is readily apparent from a comparison of Fig.~\ref{f:kmap} with
fig.~4 of Paper~I, many of these knots could be severely reddened by
their proximity to the dust lanes. We have made a first-order
correction for this in the following way. Using the isochrone
synthesis models of Bruzual \& Charlot (1993), we find that from an
initial $(I-K)$ colour of 1.48 at 20~Myr, a given population will
become bluer by $\sim0.1$~mag at 1~Gyr, then finish up redder by about
the same amount after 10~Gyr. We have therefore used the $(I-K)$ map
from Paper~I in order to calculate the colour excess (relative to
1.48) of the unresolved stellar background around each of the knots,
together with the interstellar extinction curve of Rieke \& Lebofsky
(1985), to derive appropriate corrections for reddening and extinction
at the location of each of the knots, as given in Table~\ref{t:knots}.

\begin{figure*}
\vspace{18cm}
\includegraphics{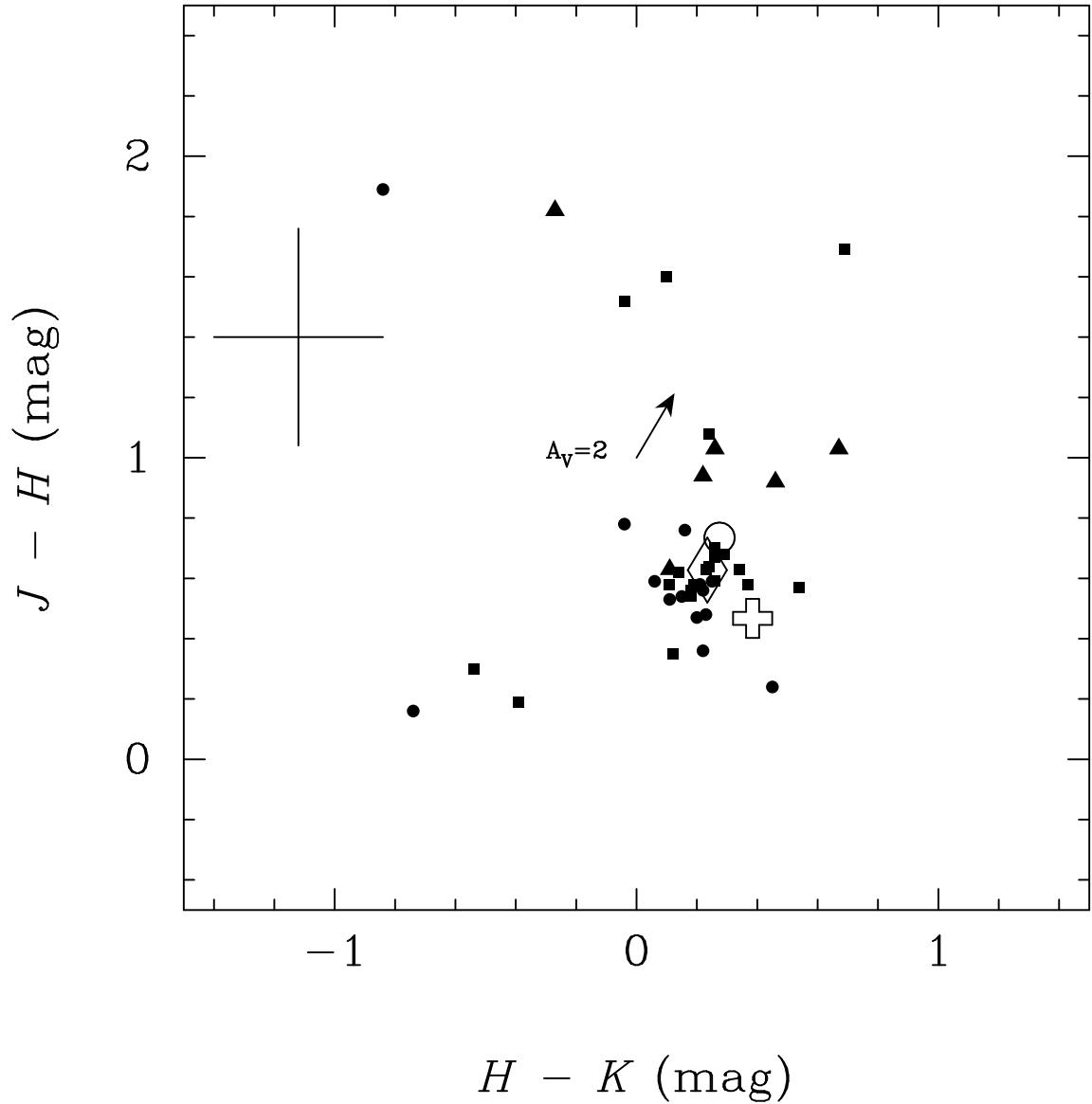}
\caption{Plot of knot NIR colours, after first-order corrections for
reddening as described in the text. The average photometric error bar
from Table~\protect{\ref{t:knots}} is shown, as is the reddening vector
for a residual visual extinction of 2~mag. The open circle marks the
`typical' colours of an old stellar population, the open diamond marks
the colours of the Leitherer \& Heckman (1995) model 300~Myr after a
burst of star formation, and the hollow cross marks the colours of
the Leitherer \& Heckman model after 300~Myr of continuous star formation.
Filled triangles are used for knots embedded within the dust lanes, while
filled squares and circles indicate knots adjacent to, or well removed from
the dust lanes, respectively.}
\label{f:twocol}
\end{figure*}

The distribution of knot NIR colours, after correcting for reddening
as above, is shown in Fig.~\ref{f:twocol}. The bulk of the knots
cluster around a point slightly bluer than the mean colours for an old
stellar population, i.e., 0.71--0.76 in $(J-H)$, and 0.25--0.30 in
$(H-K)$ (Griersmith, Hyland \& Jones 1982). Some of the knots are
redder in $(J-H)$ and bluer in $(H-K)$ than can be accounted for by
their photometric errors, or by any amount of internal reddening, warm
dust, or scattered light (see, e.g., Israel et al. 1998). These are all
diffuse, outlying regions. The three bluest knots (33, 34, and 42 in
Table~\ref{t:knots}) are all located along the inner edge of the
southwest dust lane.

\begin{figure*}
\vspace{18cm}
\includegraphics{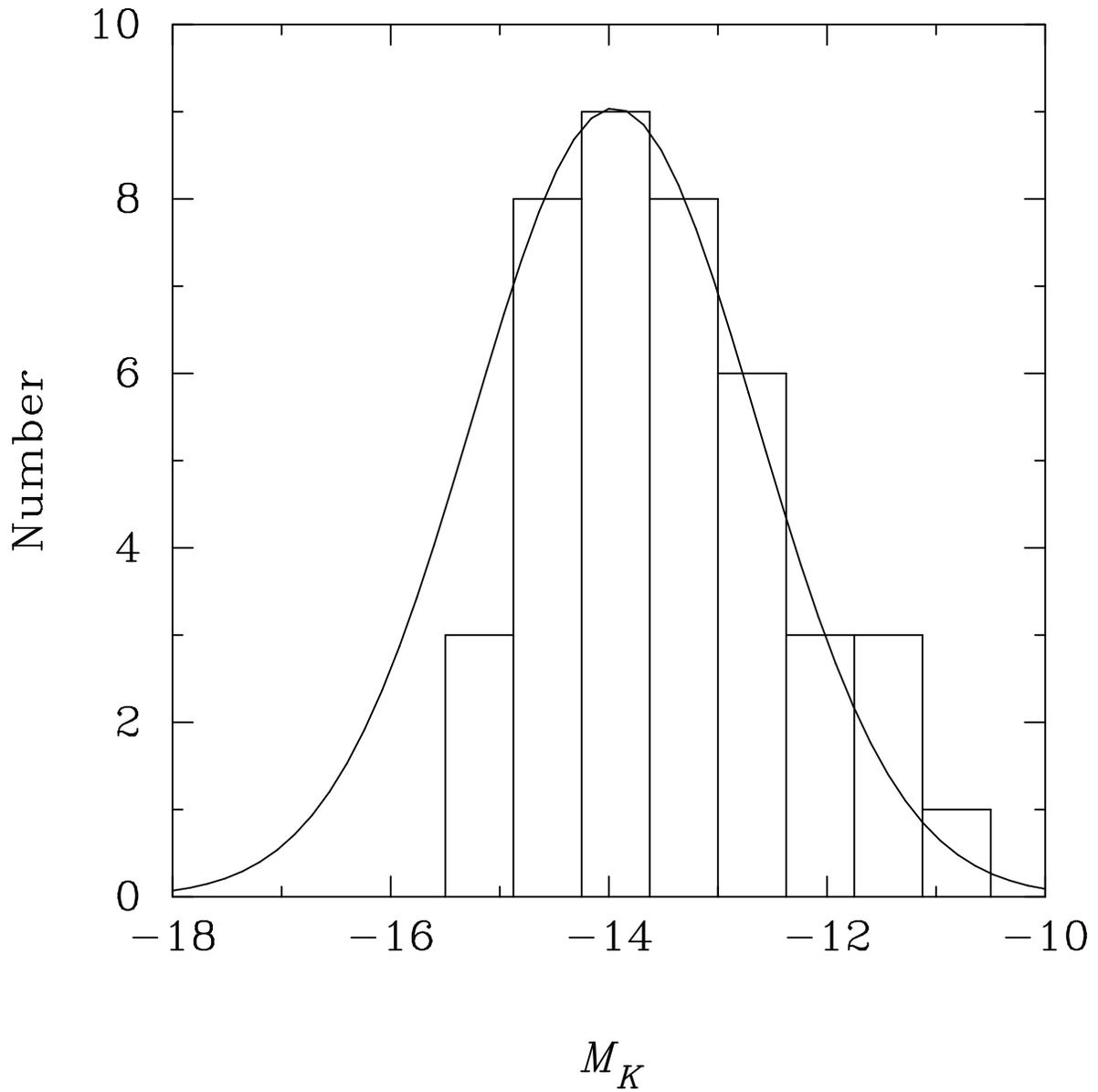}
\caption{$K$-band Luminosity Function for the 41~measurable objects in
Fig.~\protect{\ref{f:kmap}}, after corrections for extinction. We
have superimposed on this LF a Gaussian with $\langle M_K \rangle =
-13.95$, representative of globular cluster systems in nearby
galaxies, but there seems to be a deficiency of at least 6 bright
clusters in this case.}
\label{f:klf}
\end{figure*}

The $K$-band Luminosity Function (LF), corrected for extinction using
the $A_{K}$ values in Table~\ref{t:knots}, is presented in
Fig.~\ref{f:klf}.  A distance to M100 of 16.1~Mpc (Ferrarese et
al. 1996) has been assumed.  The LF peaks at $M_{K} \sim -14$, and has
a steep bright-end cutoff compared with the faint-end tail. We have
overlaid on Fig.~\ref{f:klf} the characteristic LF for extragalactic
globular cluster systems, a Gaussian having $\langle M_V \rangle =
-7.1$ and $\sigma(M_V) = 1.3$ (Harris 1991). It is possible to fit the
peak and the faint-end, provided $\langle V-K \rangle = 6.9$ (which would
require the stellar content to be totally dominated by dwarfs or
giants of type M6 or later), but there is then a clear deficiency of bright
objects if, indeed, these knots are globular clusters.


\subsection[]{Spectroscopy}\label{s:spec}

Continuum fitting and equivalent width measurements were carried out
using the {\sc splot} task in {\sc iraf}. The results are given in
Table~\ref{t:eqw}. The 1-$\sigma$~measurement errors are $\sim0.1$~nm
for the emission lines. In the case where no line feature was apparent
at the expected (redshifted by 1571~\kms) wavelength, the nearest
positive peak was measured and used as an upper limit. For the CO
measurements, we have used the wavelength range W2 of Puxley, Doyon \&
Ward (1997; hereafter PDW), equivalent to that used by Kleinmann \&
Hall (1986) and by Origlia, Moorwood \& Oliva (1993). In the last
column of Table~\ref{t:eqw}, we give the spectroscopic CO index
(CO$_{\rm sp}$) as defined by Doyon, Joseph \& Wright (1994; hereafter
DJW). The use of a broader wavelength window here is important, since
our resolution is at least a factor of 2 poorer than results published
elsewhere. K1 and K2 have quite similar CO line strengths, and both
are stronger than the nucleus.

The values observed ($\sim1$~nm) are similar to those in M83 (PDW),
and galaxies in general (Oliva et al. 1995). We note in passing that
according to the diagnostic diagram of Moorwood \& Oliva (1988), the
ratios of H$_{2}$ 1--0 S(1) to Br$\gamma$, together with the upper
limits on [Fe\,{\sc ii}], would tend to suggest that the nucleus of
M100 has some Seyfert 2 activity, while K1 and K2 would be classed as
`Composite'-like (i.e., their optical spectra would have indications
of Seyfert and H\,{\sc ii} activity). However, we feel that many of
the line detections are too marginal for the ratios to have much
meaning.

\section[]{Discussion}\label{s:disc}

\subsection[]{Old stellar clusters or recent massive star
formation?}\label{s:age}

Having resolved these compact, NIR emission knots, we now attempt to
draw some inferences about their nature. In particular, are these
knots globular clusters, dominated by an old stellar population? Or
are they much more recent, perhaps formed in a single burst, or
triggered sequentially by some dynamical process?

On the basis of the photometry alone, we might be tempted to conclude
that these knots are a few Gyr old. They are only slightly bluer than
the characteristic colours of an old stellar population.  The $K$-band
LF can be approximated by the Gaussian globular cluster LF, aside from
a deficiency of bright objects. It is the addition of the
spectroscopy, however, that provides the decisive argument in favour
of these knots being the result of recent star formation. The mere
presence of Br$\gamma$ in emission, coupled with the deep CO
absorption features attributable to young supergiants, argues strongly
that (at least for the knots associated with K1 and K2), massive star
formation has taken place within the last 100~Myr or so.

We can use our knowledge of the colours, together with the starburst
models of Leitherer \& Heckman (1995; hereafter LH95) to place some
constraints on the ages of these knots.  To begin with, we have
tracked the time evolution of $(J-H)$ and $(H-K)$ in the LH95 models
for an instantaneous burst, and for continuous star formation, using
solar metallicity, a Salpeter-type IMF, and a mass range of
1--100~M$_{\odot}$. The end-points of these two models after 300~Myr
of evolution are shown as the large diamond and cross, respectively,
in Fig.~\ref{f:twocol}. It can clearly be seen that the knot colours
on the whole are more consistent with a recent {\em burst} of star
formation, than with either continuous or ancient star formation.
However, as Satyapal et al. (1997) point out, extinction-corrected
colours alone do not provide good constraints on stellar population
models.

A more precise determination of the age of the burst requires
spectroscopic information, which is only available for the regions K1
and K2 (and in fact, the narrow slit encompasses knots 29 and 19
only). These two regions have quite similar colours, emission line
equivalent widths, and CO spectroscopic indices. Their similarity at
mid-IR wavelengths has been noted by Wozniak et al. (1998), and indeed,
the main difference between them seems to be the degree of obscuration
by dust (Paper~II). Using Br$\gamma$ equivalent widths only,
the instantaneous burst models of LH95 imply an age
of 8-13~Myr for these two regions. Even with a steeper (Miller-Scalo
type) IMF, the LH95 models with continuous star formation require
$>300$~Myr for the Br$\gamma$ emission to drop this low.

\begin{figure*}
\vspace{18cm}
\includegraphics{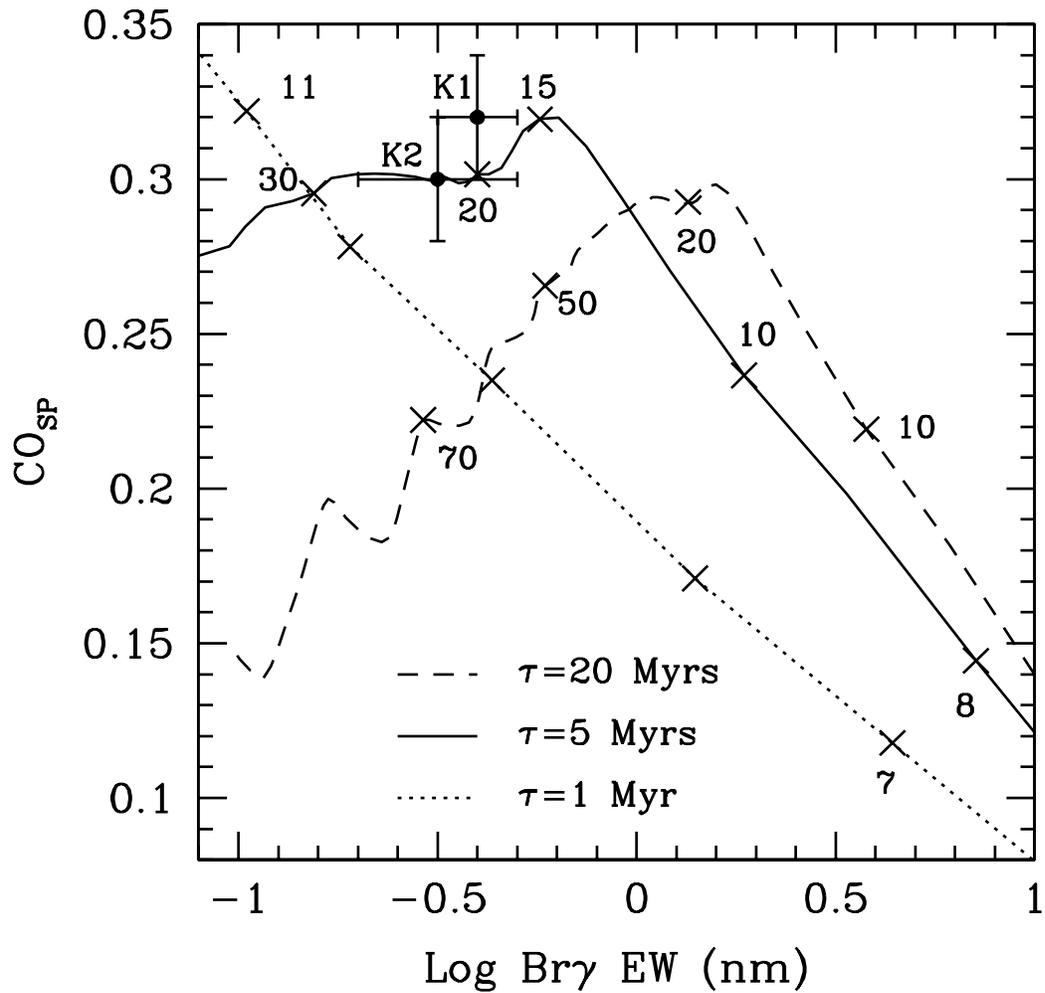}
\caption{Observed CO spectroscopic indices and Br$\gamma$ equivalent
widths for the 2 knots K1 and K2. Evolutionary tracks for starburst
models with 3~different exponential decay timescales are also shown,
reproduced from fig.~3 of Puxley, Doyon, \& Ward (1997). The numbers
on each track mark the starburst age in Myr.}
\label{f:pdwfig3}
\end{figure*}

Devost \& Origlia (1998) have combined the LH95 models with equivalent
width calculations, isochrones, and evolutionary tracks from the
literature to forecast the time evolution of the CO (6--3) 1.62~$\mu$m
and CO (2--0) 2.29~$\mu$m equivalent widths. Using their fig.~1, we
derive ages of either 5 or 15~Myr, though the weakness of Br$\gamma$
in emission makes the latter more likely. The DJW models predict the
time evolution of {\em both\/} Br$\gamma$ and CO, and by plotting CO$_{\rm sp}$
versus Br$\gamma$ line strengths in Fig.~\ref{f:pdwfig3} (adapted from
fig.~3 of PDW), we find that both K1 and K2 lie close to the locus
traced by a starburst evolutionary model having an exponential decay
time-scale of 5~Myr. The ages of K1 and K2 implied by these models are
between 15 and 25~Myr. In Paper~I, Knapen et al. reached similar
conclusions regarding the ages of the knots K1 and K2 on the basis of
$(UVK)$ colours and the models of Charlot \& Bruzual (1991).


From modelling and observations, we know that the CNR hosts two
small-scale but relatively strong density wave spiral arms (Paper~II;
Knapen 1996; Knapen et al., in preparation).  Starbursts can occur in
the spiral armlets, induced by the concentration and compression of
abundant gaseous material by the density wave amplified by the gas
self-gravity, and show up first as H\,{\sc ii}~regions, and after
$\sim5$~Myr, start emitting strongly in $K$ (LH95; Knapen 1996).  The
distribution of SF regions in the CNR is compatible with and fully
explained by the interaction of disk gas with the non-axisymmetric
(barred) potential in M100 (Paper~II).  Our results also confirm that
the individual knots of SF in the core of M100 have been created in
short 15--25 Myr bursts by the global density wave pattern. 

\subsection[]{Stellar populations in the knots}\label{s:stars}

The mean de-reddened colours of the 41 measurable knots ($\langle J-H
\rangle = 0.74$, $\langle H-K \rangle = 0.14$) are consistent with the
stellar population being dominated by giants of type K3-4, or M0
supergiants [dwarf stars being too blue in $(J-H)$]. Similarly, the
CO$_{\rm sp}$ values for K1 and K2 in Table~\ref{t:eqw} are also
typical of M0~supergiants, or of M4 giants (again, dwarfs do not have
CO$_{\rm sp}$ this high; DJW). As Oliva et al. (1995) and PDW have
shown, even the use of CO spectral features at 1.59, 1.62, and
2.29~$\mu$m is still not sufficient to distinguish giants from
supergiants. The LH95 and DJW models both predict the supergiant
production rate to peak at $\sim15$~Myr after the onset of the burst,
so a supergiant-dominated population is perhaps more likely for this
reason. In any case, we note that a single knot with $M_{K} = -14$
would require up to 25000 K4~giants, or as few as just 70
M0~supergiants, to account for the observed luminosity. Emission lines
are not expected to contribute much to the knot luminosities, as a
15~Myr old cluster will have a main sequence turn-off near spectral
type B0 (Whitmore \& Schweizer 1995), and such stars can no longer
ionise the gas around them.

A difference in the stellar populations of the knots relative to the
unresolved nuclear light also has implications for the dynamical
analyses. Our results confirm the variations in $M/L_{K}$ over the CNR
in M100 derived in Paper~II and quantified by the dynamical modelling
of Wada et al. (1998).

By comparing the $K$-band luminosities of the LH95 models 20~Myr
after a burst with Fig.~\ref{f:klf}, we estimate that between
0.1 and $4.0 \times 10^{4}$~M$_{\odot}$ of gas went into forming
each of these knots, and that the total mass of gas involved
in forming these knots was $\sim6\times10^{6}$~M$_{\odot}$. Since
most of these knots have been resolved by our observations
(${\rm FWHM}=4-7$~pixels), we are able to deconvolve the images,
and derive physical sizes for the knots of between 30 and 70~pc.

Finally, we note that these circumnuclear knots in M100 are somewhat
different from the class of `super star clusters' identified in {\em
HST\/} images of nearby starburst galaxies (e.g., O'Connell, Gallagher
\& Hunter 1994; Meurer et al. 1995), and which have been postulated to
be the progenitors of globular clusters. The knots are not only larger
in size, but also much redder than the `super star clusters', even
though both types of object are thought to be $\sim15$~Myr old. Since
the nucleus of M100 is not currently undergoing an intense starburst,
these differences could reflect the different types of environment in
which they formed.

\section{Conclusions}\label{s:conc}

Recent improvements in the delivered image quality at {\em UKIRT\/} have
allowed us to resolve a number of compact circumnuclear `knots' in M100. 
NIR colours and luminosities, corrected for extinction, are consistent
with a recent burst (or bursts) of massive star formation as the origin
of these knots, but the possibility of an older population of objects
(e.g., globular clusters) cannot be ruled out using this data alone. 
The addition of NIR spectroscopy for two of these knots either side of
the nucleus provides conclusive evidence of recent star formation.  By
comparing Br$\gamma$ and CO 2.29~$\mu$m line equivalent widths to
various starburst models in the literature, we infer ages for these
knots of between 15 and 25~Myr, and a stellar population most likely
dominated by M0~supergiants.  Future NIR spectroscopy for the remainder
of the knots should enable us to generalize this conclusion and refine
our interpretation of the formation and evolution of the SF regions in
the presence of the small-scale spiral density wave. 

\section*{Acknowledgments}

We are grateful to Tom Geballe and Sandy Leggett for generously making
time available on {\em UKIRT\/} for this project, to Ken Mighell for
offering us the use of his software, and to Ren\'{e} Doyon for allowing us
to make use of the models and figure from the PDW paper. We are indebted
to Isaac Shlosman, Tim Hawarden, and the referee for constructive
criticism on an earlier version of the manuscript.

\bsp


\begin{table*}
 \caption{Derived Parameters for Compact Knots in M100}
 \label{t:knots}
 \begin{tabular}{rrrccccc}
\hline
No. & \multicolumn{1}{c}{$\Delta\alpha$} & \multicolumn{1}{c}{$\Delta\delta$} &
  $K~(\pm)$  & $(J-H)~(\pm)$ &
$(H-K)~(\pm)$ & $E(I-K)$ & $A_{K}$ \\
    & (arcsec)$^{a}$ & (arcsec)$^{a}$ &             &               &
              &          &         \\
\hline
 1  &  -6.3 &  12.6 & 18.47 (0.10) & 2.16 (0.95) & -0.07 (0.19) & 1.18 & 0.35\\
 2  &  -4.1 &  12.3 & 16.13 (0.03) & 1.43 (0.12) &  0.50 (0.07) & 1.39 & 0.42\\
 3  &   0.3 &  12.9 & 19.32 (0.22) & 1.98 (0.90) & -0.79 (0.34) & 0.30 & 0.09\\
 4  &   2.8 &  12.6 & 16.85 (0.05) & 0.97 (0.11) &  0.07 (0.09) & 0.66 & 0.20\\
 5  &   6.2 &   7.0 & 16.95 (0.08) & 0.75 (0.16) &  0.15 (0.15) & 0.57 & 0.17\\
 6  &   5.0 &   6.6 & 17.40 (0.14) & 1.85 (1.03) &  0.78 (0.36) & 0.55 & 0.17\\
 7  &  -1.2 &   8.0 & 17.16 (0.14) & 0.37 (0.24) &  0.22 (0.28) & 0.03 & 0.01\\
 8  &  -0.7 &   7.1 & 17.22 (0.16) & 0.25 (0.33) &  0.45 (0.36) & 0.03 & 0.01\\
 9  &   2.4 &   6.0 & 18.62 (0.80) & 1.08 (2.38) &  0.36 (1.75) & 0.71 & 0.21\\
10  &   3.8 &   4.6 & 16.63 (0.10) & 0.56 (0.19) &  0.24 (0.20) & 0.36 & 0.11\\
11  &   5.1 &   3.6 & 16.26 (0.08) & 1.36 (0.35) &  0.87 (0.21) & 1.15 & 0.35\\
12  &   4.2 &   3.5 & 16.98 (0.16) & 0.50 (0.26) &  0.21 (0.30) & 0.53 & 0.16\\
13  &   7.6 &   1.3 & 16.34 (0.09) & 0.63 (0.15) &  0.23 (0.17) & 0.30 & 0.09\\
14  &   9.2 &   0.7 & 16.88 (0.12) & 0.92 (0.23) &  0.26 (0.23) & 0.57 & 0.17\\
15  &   8.2 &   0.2 & 16.09 (0.07) & 0.60 (0.13) &  0.30 (0.14) & 0.42 & 0.13\\
16  &   7.4 &  -0.4 & 16.31 (0.10) & 0.71 (0.19) &  0.33 (0.20) & 0.43 & 0.13\\
17  &   7.8 &  -1.4 & 16.29 (0.09) & 0.64 (0.15) &  0.17 (0.17) & 0.38 & 0.11\\
18  &   7.5 &  -2.2 & 15.88 (0.07) & 0.66 (0.12) &  0.28 (0.13) & 0.34 & 0.10\\
19  &   6.8 &  -3.2 & 15.47 (0.05) & 0.70 (0.10) &  0.38 (0.10) & 0.25 & 0.08\\
20  &   5.9 &  -3.7 & 15.31 (0.04) & 1.15 (0.13) &  0.59 (0.09) & 0.79 & 0.24\\
21  &   5.8 &  -6.0 & 16.00 (0.06) & 0.77 (0.13) &  0.32 (0.12) & 0.67 & 0.20\\
22$^{b}$  &   5.4 &   0.0 & $\ldots$ & $\ldots$ & $\ldots$ & 0.81 & 0.24\\
23  &   4.5 &  -1.5 & 15.66 (0.06) & 0.89 (0.12) &  0.26 (0.11) & 0.90 & 0.27\\
24  &   3.8 &  -0.5 & 15.48 (0.05) & 0.74 (0.11) &  0.21 (0.10) & 0.41 & 0.12\\
25  &   3.3 &   1.2 & 17.19 (0.31) & 1.72 (1.39) &  0.17 (0.60) & 0.42 & 0.13\\
26  &  -0.4 &   4.0 & 16.23 (0.09) & 0.64 (0.16) &  0.09 (0.17) & 0.16 & 0.03\\
27  &  -4.3 &   2.1 & 16.24 (0.10) & 1.15 (0.28) &  0.35 (0.20) & 0.74 & 0.22\\
28  &  -6.5 &   4.6 & 15.39 (0.04) & 0.66 (0.08) &  0.22 (0.08) & 0.40 & 0.12\\
29  &  -6.4 &   3.4 & 15.14 (0.03) & 0.74 (0.07) &  0.29 (0.06) & 0.37 & 0.11\\
30  &  -7.8 &   2.4 & 14.87 (0.03) & 0.87 (0.07) &  0.40 (0.06) & 0.66 & 0.20\\
31  &  -8.5 &   0.5 & 15.65 (0.05) & 0.79 (0.11) &  0.31 (0.10) & 0.32 & 0.10\\
32$^{b}$  &  -6.8 &  -0.1 & $\ldots$ & $\ldots$ & $\ldots$ & 0.88 & 0.26\\
33  &  -5.8 &  -0.6 & 18.43 (0.88) & 0.42 (0.66) & -0.47 (1.23) & 0.40 & 0.12\\
34  &  -5.4 &   0.2 & 17.95 (0.54) & 0.25 (0.33) & -0.69 (0.73) & 0.31 & 0.09\\
35  &  -5.0 &  -3.7 & 15.47 (0.04) & 0.77 (0.11) &  0.32 (0.09) & 0.34 & 0.10\\
36  &  -0.4 &  -5.4 & 17.00 (0.15) & 0.66 (0.28) &  0.16 (0.28) & 0.28 & 0.08\\
37  &   0.3 &  -6.4 & 16.33 (0.08) & 0.53 (0.14) &  0.23 (0.15) & 0.20 & 0.06\\
38  &  -3.8 &  -7.0 & 17.42 (0.17) & 0.68 (0.53) &  0.60 (0.41) & 0.38 & 0.11\\
39  &  -2.5 &  -8.4 & 16.52 (0.07) & 0.64 (0.15) &  0.24 (0.14) & 0.01 & 0.00\\
40  &  -1.7 &  -9.1 & 17.08 (0.10) & 0.68 (0.27) &  0.43 (0.22) & 0.36 & 0.13\\
41  &   0.1 &  -9.1 & 18.25 (0.33) & 0.65 (0.72) &  0.23 (0.65) & 0.25 & 0.08\\
42  &   1.0 & -10.2 & 18.29 (0.26) & 0.19 (0.25) & -0.39 (0.40) & 0.00 & 0.00\\
43  &  -0.3 & -11.5 & 17.62 (0.09) & 1.76 (0.42) &  0.10 (0.18) & 0.82 & 0.25\\
\hline
\end{tabular}
\medskip
~~\\
\begin{flushleft}
$^{a}$Offset relative to $K$-band nuclear position.\\
$^{b}$Insufficient signal-to-noise for centering and photometry. \\
\end{flushleft}
\end{table*}

\begin{table*}
 \caption{Line Equivalent Widths in Selected Regions of M100}
 \label{t:eqw}
 \begin{tabular}{lcccccc}
\hline
Region  & --EW([Fe\,{\sc ii}]) & --EW(H$_{2}$ 1--0 S(1)) & --EW(Br$\gamma$) &
--EW(H$_{2}$ 2--1 S(1)) & EW(CO 2--0) & CO$_{\rm sp}$ \\
        &     (nm)                      &        (nm)                      &
 (nm)                      &     (nm)                         &
 (nm)                  &               \\
\hline
Nucleus & $<0.1$                        & 0.3                              &
$<0.3$                     & $<0.2$                           &
0.88                   &  $0.24\pm0.01$ \\
K1      & $<0.07$                       & 0.3                              &
0.4                        & $<0.2$                           &
1.01                   &  $0.32\pm0.02$  \\
K2      & $0.2$                         & 0.2                              &
0.3                        & 0.1                              &
0.93                   &  $0.30\pm0.02$  \\
\hline
\end{tabular}
\end{table*}

\label{lastpage}

\end{document}